\documentstyle[11pt,psfig]{article}
\begin{document}
\vskip 0.2in
\begin{center}
\vskip 0.1in
\Large{\bf
SINGLE PHOTONS}\\
\vskip 0.1in
\Large{\bf FROM}\\
\vskip 0.1in
\Large{\bf RELATIVISTIC HEAVY ION COLLISIONS}\\
\vskip 0.1in
\Large{\bf AND}\\
\vskip 0.1in
\Large{\bf QUARK-HADRON PHASE TRANSITION}\\
\vskip 0.2in
\large{\bf 
Dinesh Kumar Srivastava}
\vskip 0.2in
\it {Variable Energy Cyclotron Centre,\\
     1/AF Bidhan Nagar, Calcutta 700 064,\\
     India}
\end{center}

\renewcommand{\baselinestretch}{1.5} 
\parindent=20pt

\vskip 0.1in
\begin{center}
\bf {Abstract}\\
\end{center}
The present status of theoretical expectations of studies of
single photons from relativistic heavy ion collisions is discussed.
It is argued that the upper limit of single photon radiation
 from S$+$Au collisions at CERN
SPS obtained by the WA80 collaboration perhaps rules out any reasonable
description of the collision process which does not involve a phase
transition to quark gluon plasma. Predictions for 
single photons from the quark-matter likely to be created
in collision of two lead
nuclei at RHIC and LHC energies are given with a proper accounting of
chemical equilibration and transverse expansion. Finally, it is pointed out
that, contrary to the popular belief of a quadrilateral dependence of
electromagnetic radiations ($N_\gamma$) from such collisions on the 
number of charged particles ($N_{\mathrm {ch}})$, we may only have
$N_\gamma \, \propto \,N_{\mathrm {ch}}^{1.2}$. 

\vskip 0.25in
\section{INTRODUCTION}

Relativistic heavy ion collisions are expected to lead to
observation of a QCD phase transition from hadronic matter to
quark matter and an ephemeral formation of Quark
Gluon Plasma (QGP). 
Single photons and  dileptons have long been considered as excellent probes of
the early stages of relativistic heavy ion collisions. Their usefulness
stems from the fact that - once produced - they hardly ever interact and
leave the system with their energy and momentum unaltered. Their production
cross-section is also known to increase rapidly with temperature and they
should provide valuable information about the hot and dense - the truly exotic-
stage of the matter likely to be created in such collisions.
These early expectations have been
considerably refined in recent times. It is now realized
that we have to obtain a quantitative understanding of the 
various sources of single photons and dileptons before we are ready
for an experimental identification of the thermal radiations from QGP.
By now, it has also become evident that - at least - at the SPS energies,
the initial temperatures are not likely to be very high.  
Thermal radiations are emitted at every stage of the evolution of the 
interacting system. Thus for an expanding system, the
radiations from the late hadronic phase can easily overwhelm the
radiations from the initial QGP phase if the initial temperatures are
not very high. This will be more so at lower transverse momenta.

The experimental identification of single photons gets very difficult
due to the huge background of photons originating from the decay of
$\pi^0$ and $\eta$ mesons. While there is some hope that
it may be possible to isolate single photons at RHIC and LHC energies
on an event by event basis; at the SPS energies this separation can only
be attempted on a statistical basis. 

 The only guiding factor at the moment is the upper limit for single
 photons from S$+$Au collisions at SPS energies obtained recently by
 the WA80 collaboration~\cite{terry}. 
 The situation is quite lively, on the other
 hand, for the case of lepton pairs as a number of experiments have
 reported  excess production of lepton pairs,
 both from sulphur and lead induced collisions~\cite{axel} at the
 same energies.
 This already  provides for a very important conclusion, which is often
 missed, viz.; the mechanism for the production of dileptons and single
 photons being similar, if there is an excess production of dileptons, there
 must be a  production of single photons. We have to assess the experimental    situation against the backdrop of these considerations. 
 
  The theoretical understanding of the production of single photons has
  three essential inputs, the rate of production of single photons from
  quark matter, the rate of production of single photons from hadronic
  matter, and the mechanism for the evolution of the interacting system.
  Recent years have witnessed enormous developments on all these fronts.
   Thus, the rate of production of single photons from quark matter
   has been calculated by a number of authors; with addition of soft
   and hard contribution using the Bratten - Pisarski resummation method
   to shield the singularity for a baryon free
   plasma~\cite{joe,rud1}, for a plasma with a finite baryo-chemical
   potential~\cite{traxler}, and for a non-equilibrated plasma~\cite{rud2}
   (see also~\cite{henning,hirano,roy, str} for additional work in this
   connection). The rate of production of single photons from a hot hadronic
   matter due to a host of reactions has been studied by a number of
   authors~\cite{joe,roy,li,song,kevin,kim,steel}.  The early works~\cite{sh}
   on estimating single photons used the boost-invariant
   hydrodynamics of Bjorken~\cite{bj} without transverse expansion extensively.
   The other extremes of Landau hydrodynamics~\cite{land} and a free
   streaming expansion~\cite{kms,nad} have also been used. More recent
   studies have properly accounted for the transverse expansion of the 
   plasma~\cite{prd,arbex,dum,neu,lok,yu,soll,crs1}. Radiation of thermal
   photons from chemically equilibrating plasma has also been evaluated
   recently~\cite{str,burk,sgm}.
   
    We shall not attempt a review of these
   developments. Instead, as indicated earlier, we shall try to use
    the upper limit of single photons seen by the WA80 collaboration
    to constrain the theoretical description attempted in the
   literature. First, we  briefly discuss the treatment of Srivastava
    and Sinha~\cite{prl}. We discuss the assumptions 
    and give predictions for
   Pb$+$Pb collisions at SPS energies from the recent work of Cleymans,
   Redlich, and Srivastava~\cite{crs1}.
    Then we summarize our conclusions for SPS
   energies. Next we give the predictions for RHIC and LHC energies for
   single photons from a chemically equilibrating QGP~\cite{sgm}.
   Finally we describe
   a scaling behaviour between the number of single photons and the
   number of charged particles, seen from our analysis~\cite{crs2}.

\begin{figure}[t]
\centerline{\psfig{figure=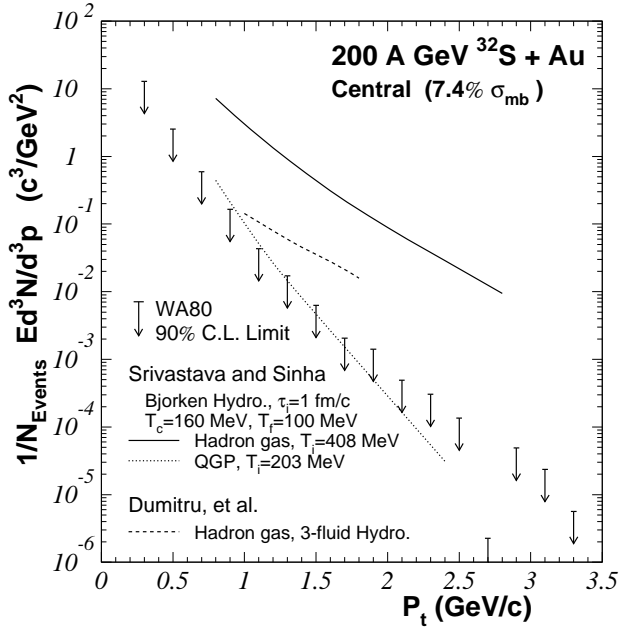,height=90mm}}
{\caption{ 
A comparison of theoretical predictions of Srivastava and Sinha with the
upper limit of single photons from central collisions of S$+$Au
 system obtained by WA80 collaboration.  The no-phase transition
 scenario, when the resulting hot hadronic gas is assumed to consist
 of $\pi$, $\rho$, $\omega$, and $\eta$ mesons, is seen to clearly 
 excluded by the upper limits of the data}}

\label{fig:sau}
%
\hspace{\fill}
\end{figure}
   
\section{RESULTS FOR SPS ENERGIES}

\subsection{S$+$Au Collisions}
Consider~\cite{prl} central collisions of the S$+$Au system 
and assume an interacting
system having an initial transverse radius $R_T$ equal to the radius of the
sulphur  nucleus and  $dN_{\mathrm {charge}}/dy=150$ at $y=0$. Assuming an
isentropic expansion one can relate the  particle rapidity density
$dN/dy\,\approx\,1.5\times dN_{\mathrm {charge}}/dy$,
the initial temperature ($T_i$), and the initial time ($\tau_i$) 
as~\cite{HK},
\begin{equation}
T_i^3\tau_i=\frac{2\pi^4}{45\zeta(3)\,\pi R_T^2\,4 a_k}\,\frac{dN}{dy}
\end{equation}
where $a_k=a_Q=37\pi^2/90$ if the system is initially in the QGP phase, 
consisting
of (massless) `$u$' and `$d$' quarks, and gluons. If, however, the system is
initially in a hadronic phase, consisting of $\pi$,~$\rho$,~$\omega$, and
$\eta$ mesons, we have $a_k=a_H\approx 4.6\pi^2/90$
appropriate for temperatures in the range 100--400 MeV. For the initial
time $\tau_i$ we take the canonical value of 1 fm/c.

\begin{figure}[t]
%
\centerline{\psfig{figure=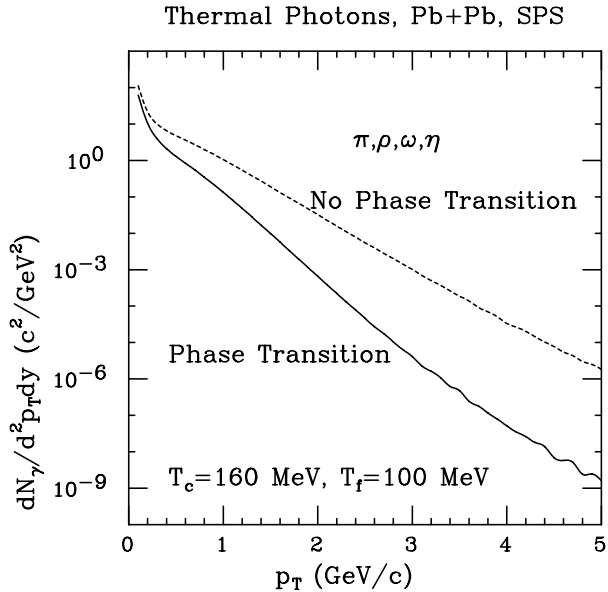,height=9cm}}
{\caption{ Transverse momentum distribution of photons at $y=0$ 
produced from the collision of lead nuclei at CERN SPS.
The hadronic matter is assumed to consist of  $\pi$, $\rho$, 
$\omega$, and $\eta$ mesons with (dashed lines)
 and without (full lines) phase transition}}
\label{fig:pb12}
\end{figure}

Thus, if the system is formed in the QGP phase at $\tau_i=$ 1 fm/c, we have
$T_i=$ 203.4 MeV. 
On the other hand,  if we assume the system to be produced in the hot hadronic
phase, with the same entropy density as before at $\tau_i=$ 1 fm/c, we get
$T_i=$ 407.8 MeV.
Thus we consider two different scenarios. In the {\em first} scenario
we assume the matter to be formed in a QGP phase at the initial  time
$\tau_i$ and initial temperature $T_i$, which then expands and
cools, and goes into the mixed phase at the transition temperature $T=T_c$.
When all of the quark matter has adiabatically converted into 
 hadronic matter,
it  cools again and undergoes a freeze-out at $T=T_f$.

In the {\em second} scenario
we consider the system to be formed in a hadronic phase with the same
entropy density as before,
at the initial time $\tau_i$ and an initial temperature $T_i$ which
expands, cools, and undergoes a freeze-out at $T_f$, {\em without admitting a
QCD phase transition}.

We assume a boost-invariant hydrodynamic expansion with transverse 
expansion of the system (Ref. ~\cite{prd,vesa}).
The thermal photon spectrum is obtained by convoluting the rate of emission
of photons with the space-time evolution of the system, using methods
which are well established by now ~\cite{prd}.
For emission of photons from
the QGP we consider the Compton plus annihilation contribution corrected
for infrared divergences as ~\cite{joe},
\begin{equation}
E\frac{dR}{d^3p}= \frac{5}{9}\,\frac{\alpha \alpha_s}{2\pi^2}\,T^2
\,e^{-E/T}\,\ln\left[1+\frac{2.912\,E}{g^2\,T}\right]
\end{equation}
where $\alpha_s$ is the strong coupling constant. For the
hadronic matter we explicitly consider all the reactions involving the
complete list of (nonstrange) light mesons
($\pi\pi\rightarrow\rho\gamma,\, \pi\rho\rightarrow\pi\gamma,
\,\pi\pi\rightarrow\eta\gamma,\,\pi\eta\rightarrow\pi\gamma,\,$ and $\pi^+
\pi^-\rightarrow\gamma\gamma$)
and the decay of vector mesons ($\omega^0\rightarrow\pi^0\gamma$ and
$\rho^0\rightarrow\pi^+\pi^-\gamma$)
considered  by Kapusta, Lichard and Seibert ~\cite{joe}.
 In addition we include the contribution of
$\pi\,\rho\,\rightarrow\, A_1\,\gamma$ through the
parametrization suggested by Xiong et al.~\cite{li} whose results are
rather similar to those of Song~\cite{song}.

The photon spectrum is then obtained as,
\begin{equation}
\frac{dN}{d^2p_T\,dy}=\int\left[f_Q(r,\tau,\eta)
\left(E\frac{dR}{d^3p}\right)_{\mathrm {QGP}}
+f_H(r,\tau,\eta)\left(E\frac{dR}{d^3p}\right)_{\mathrm {Had}}\right]\tau\,d\tau\,
r\,dr\,d\eta\,d\phi~,
\end{equation}
where $f_Q$ is the fraction of the quark-matter in the system and $f_H$ is
the hadronic fraction. 
 We take $T_c=$ 160 MeV and assume the freeze-out to take place
at 100 MeV. In any case the thermal photon production becomes insignificant
at lower temperatures.

The results of this  study are compared with the upper limit of the
data seen by the WA80 group in Fig.~1 (Ref.~\cite{terry}).

It is seen that a hadronic gas description is clearly ruled out by the
upper limit of the data. Similar conclusions were reached by a 
number of authors~\cite{arbex,dum,neu,soll} using widely different
descriptions for the expansion of the system. However, 
it should be emphasized
that an initial temperature of about 400 MeV for the hot hadronic gas
is already unacceptable from simple physical considerations as it would
amount to a hadronic density in excess of 10 hadrons/fm$^3$.

\begin{figure}[t]
%
\centerline{\psfig{figure=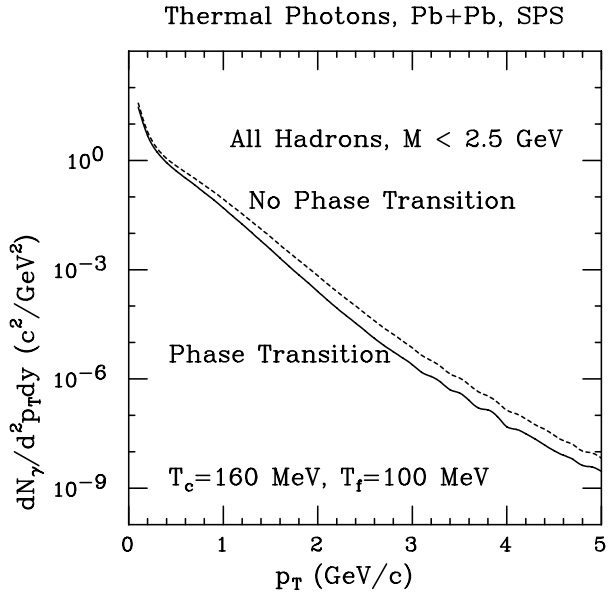,height=9cm}}
{\caption{ Transverse momentum distribution of photons at $y=0$ 
produced from the collision of lead nuclei at CERN SPS.
The hadronic matter is assumed to consist of  all hadrons in particle data
book, with (dashed lines) and without (full lines) phase transition}}

\label{fig:pb34}
\end{figure}

\subsection{Predictions for Pb$+$Pb collisions at SPS}

These conclusions on the basis of sulphur induced collisions have been
closely scrutinized. Thus, it was argued that, one could consider a
much richer constitution of the hadronic matter instead of limiting ourselves
to only light mesons. This would bring down the initial temperature.
However, there is still no published estimate of the rates for emission
of photons from hadronic reactions which might involve heavier mesons
and, say,  baryons. One may still obtain a lower limit of hadronic
description for the data by considering a much richer hadronic matter
for the equation of state, but still employing the rates used earlier in
Ref.~\cite{prl}. Sollfrank et al.~\cite{soll} have reported predictions
in agreement with the upper limit of the WA80 results using such
a procedure. However, one
must remember that the initial energy density for this particular
description  in the above work~\cite{soll} is
 several GeV/fm$^3$ and the hadronic density is also several hadron/fm$^3$,
 which does not inspire confidence in a hadronic description.
 
\begin{figure}[t]
%
\centerline{\psfig{figure=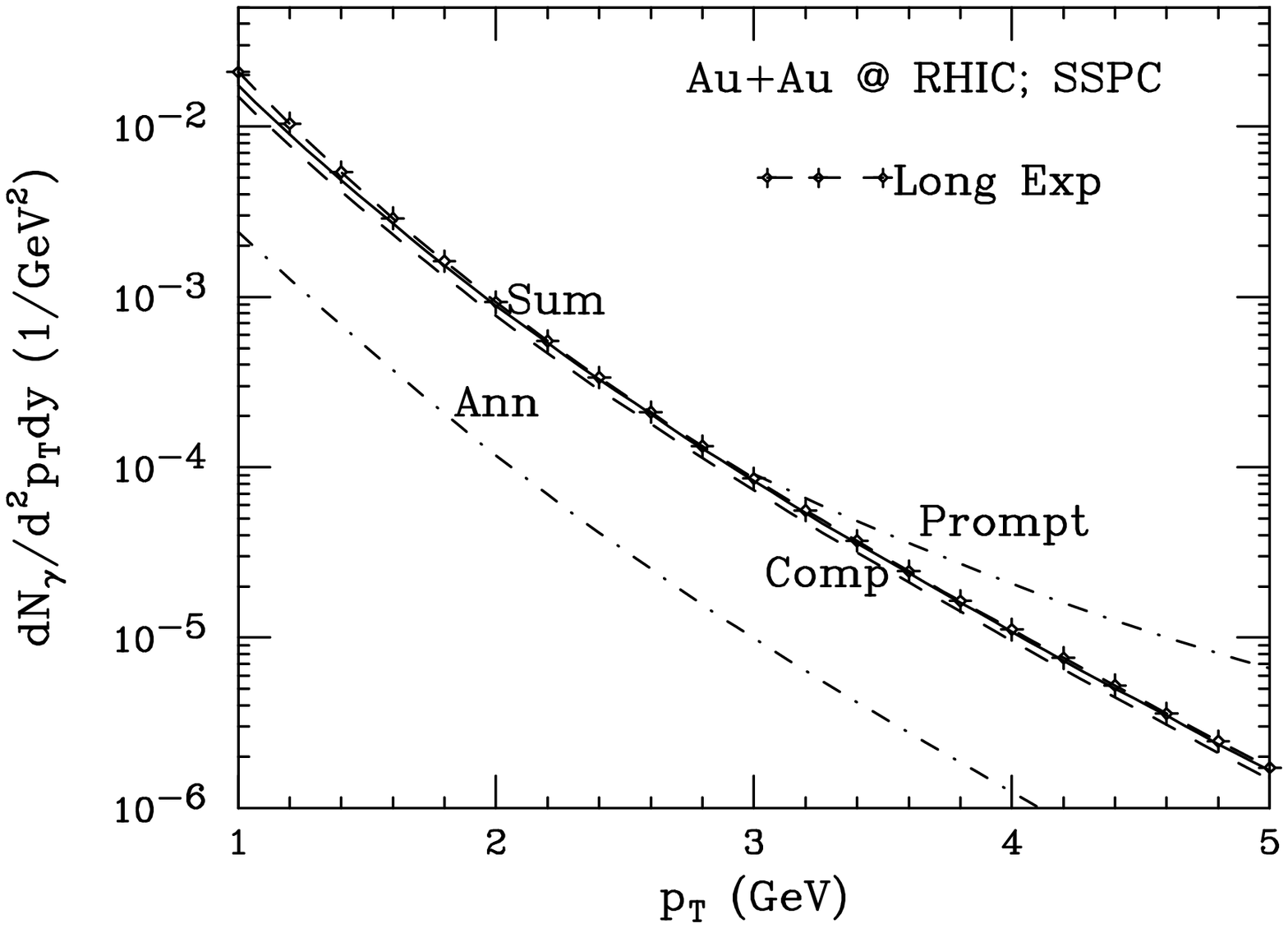,height=9cm}}
{\caption{ Distribution of thermal photons from the QGP phase at RHIC,
from a chemically equilibrating and transversely expanding plasma.
The initial conditions were obtained from a self screened parton
cascade model. Results are also given for a purely longitudinal flow.
Prompt photons, whose production is governed by structure functions
are seen to dominate the yield for $p_T \, > \,$ 3--4 GeV.}}

\label{fig:rhic}
\end{figure}

  A good understanding of the effect of the equation of state of hadronic
  matter on the single photons at SPS energies is obtained from the
  detailed study of Cleymans, Redlich, and Srivastava~\cite{crs1}. Two 
  extreme prescriptions for the hadronic matter were used. In the
  first case, the hadronic matter was assumed to consist of only $\pi$,
  $\rho$, $\omega$, and $\eta$ mesons, while in the other case all
  hadrons from the particle data book were assumed to populate the 
  system in complete thermal and chemical equilibrium. Justification
  for the chemical equilibrium, at least at the time of freeze-out,
  is available from the studies of Braun-Munzinger et al.~\cite{johanna}.
  Both hadronic matter descriptions were employed with and without
  phase transition. In each case, the initial temperature was fixed
  by requiring that $dN_{\mathrm {charge}}/dy \approx 550$ 
  in collisions involving two lead nuclei.
   
  As expected from the earlier findings of Srivastava and Sinha~\cite{prl},
  the results for particles as well as photons were quite different
  for the equation of state employing a limited number of light mesons
   for the cases  with and without phase transition.  
  On the other hand, however, the predictions for photons
  as well as particle distributions were seen to be quite similar
  for the two scenarios - with and without phase transition - when the
  hadronic matter was assumed to consist of all hadrons 
  (see, e.g., Figs.~ 2 \& 3).

Before drawing any conclusion from Fig.~3 about the suitability of
single photons for distinguishing between the scenarios which may or
may not involve a phase transition, we should remember that even
though  the initial temperature in the hadronic gas scenario here
is only about 210 MeV,  which by itself may not be very large, the
initial number density is almost 3--4 hadrons/fm$^3$. This is obviously
too large, for the hadronic description to be taken seriously. One may
argue that all the hadrons may not be in chemical equilibrium at the
initial time. Then, it is quite likely that the initial temperature
would be much higher and lead to a much higher production of single photons.
Again, we insist that, as only the limited number of
hadronic reactions  evaluated
in  Refs.~\cite{joe,li} were included in these analyses,
 the no-phase-transition scenario results are only a lower bound of the
 expected results. Finally, we add that it is often argued that, the
 initial time of 1 fm/$c$ assumed in these analyses is perhaps too small.
 Increasing it to 2 fm/$c$ was found to have negligible effect on the
 QGP scenario, due to very small space-time occupied by the QGP
 phase at the SPS energies, at not too large values of the transverse
 momenta.  The hadronic gas description is affected more strongly, but
 the initial number density still remains too large ( see Ref.~\cite{ash},
also for a discussions of effect of viscosity).
 
 We may conclude therefore that the upper limit of the single photons obtained
 by the WA80 experiment 
 perhaps rules out a reasonable description of the collision
 which does not involve a phase transition. We eagerly await the
 results from the Pb$+$Pb experiment. In any case, the two predictions
 given in Fig.~3 differ by a factor of 2, and hence a data accurate to
 better than that will hopefully be able to clearly distinguish between the
 two descriptions.

We may add that the same approach is able to describe~\cite{ssg} the
excess production of low mass dielectrons (near and beyond the $\rho$ peak) 
seen in $S+Au$ collisions by
the CERES experiment~\cite{axel} without any free parameters; which 
further enhances  our confidence in the observation that we are perhaps
witnessing the quark-hadron phase transition.

\begin{figure}[t]
%
\centerline{\psfig{figure=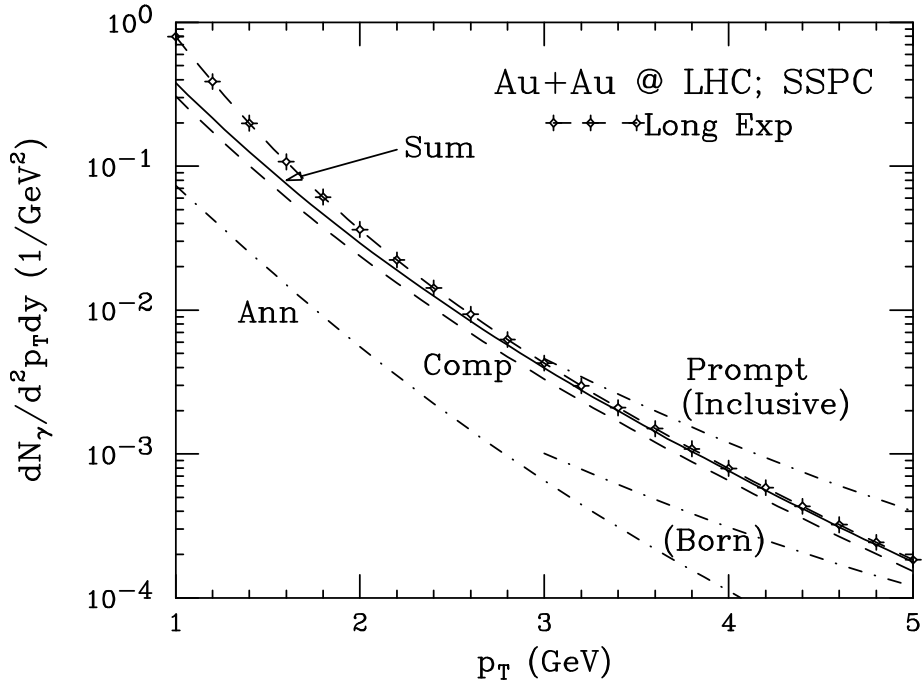,height=9cm}}
{\caption{ Same as Fig.~4 at LHC. The prompt photons from fragmentation
of quark jets are also shown.}}

\label{fig:lhc}
\end{figure}

\section{RESULTS FOR RHIC \& LHC ENERGIES}

It is generally believed that the formation of QGP at RHIC and LHC
energies in collisions involving heavy nuclei is perhaps beyond doubt. 
The formation, thermalization, and chemical equilibration of the quark
matter produced at these energies has been a subject of intense study
during the last several years. Recently the initial conditions likely to
be attained in such collisions have been obtained in a self-screened parton
cascade model~\cite{sspc}. It has been found that while a thermalization
of the plasma is obtained quickly, (say by $\tau=$ 0.25 fm/$c$), the plasma
is far from chemical equilibrium. The chemical equilibrium is likely
to proceed via gluon multiplication $(gg \leftrightarrow ggg)$ and
quark production $(gg \leftrightarrow q\overline{q})$. The transverse 
expansion of the plasma has been found to impede the chemical equilibration
due to enhanced cooling of the partonic matter. Single photons
and dileptons will prove to be invaluable tools for probing the early stages 
of such matter. We give the predictions for production of thermal
photons from such matter in Figs.~4 \& 5, and invite the reader to
Ref.~\cite{sgm} for details. 

We would like to add that at these energies, the contribution to single
photons from the plasma comes mainly from the Compton scattering of 
quarks and gluons, as the plasma is gluon-rich and quark-poor. This brings 
about an interesting and unique possibility of obtaining information about
the partonic distribution at very early times from a comparison
of single photon and dilepton measurements, as the latter will be
fully determined by the density of the quarks.

\begin{figure}[t]
%
\centerline{\psfig{figure=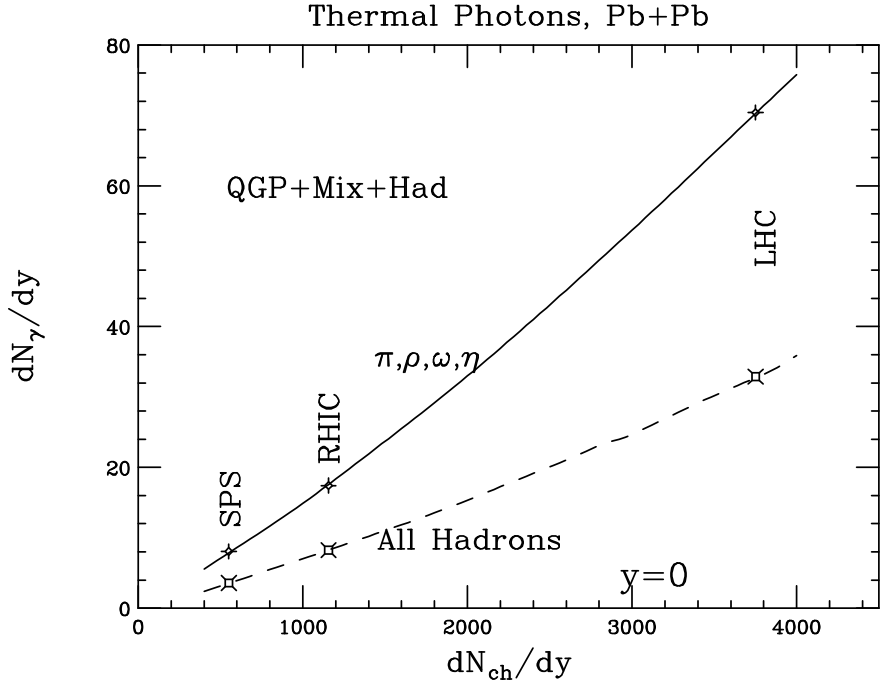,height=9cm}}
{\caption{ Variation of rapidity density of thermal photons with the
charged particle rapidity density with the two equations of state
in Fig.~3 \& 4, involving phase transition. $N_\gamma$ is seen to scale
as $K N_{\mathrm {ch}}^\alpha$ with $\alpha \approx 1.2$. $K$ is decided
by equation of state.}}

\label{fig:scaling}
\end{figure}

\section{THE SCALING $N_\gamma \propto N_{\mathrm {ch}}^\alpha$}

It is popularly believed that the number of photons (real or virtual)
radiated from relativistic heavy ion collisions, $N_\gamma$, should scale 
with the number of charged particles, $N_{\mathrm {ch}}$,
as $N_\gamma \propto N_{\mathrm {ch}}^2$. It is important to examine this
scaling behaviour as it is used to estimate the extent of the signal of
single photons against the back-ground of decay photons and even to 
figure out whether such a signal does exist at all. The two hadronic
equations of state with a provision for QGP phase transition, used
in Figs.~2 \& 3, were employed to estimate single photons for a number
of charged particle multiplicity densities~\cite{crs2}. 
As seen from Fig.~6, the
number of photons scales as $N_{\mathrm {ch}}^{1.2}$, with the constant
of proportionality decided by the equation of state. In, fact it is
rather easy to understand this as follows.  Consider a system consisting
of $N_{\mathrm {ch}}$ charged particles. The number of thermal
photons $N_\gamma$ will be given by
\begin{equation}
N_\gamma \sim e^2 N_{\mathrm {ch}} \nu
\end{equation}
where $\nu$ is the number of collisions that each particle suffers.
If the system lives long enough, as when it is confined in
a box, every particle will have a chance to collide with every other 
particle, and $\nu \sim N_{\mathrm {ch}}$. This will lead to the
quadratic dependence suggested earlier. However, the number of collisions
suffered by the particles will be given by $R/\lambda$, where $R$ is
the size and $\lambda$ is the mean free path of the particles, for a
system created in heavy ion collisions. Realizing that the number of
particles will scale as $R^3$, we get a scaling behaviour as 
$N_{\mathrm {ch}}^{4/3}$ which is quite similar to the behaviour seen here.
We should also add that in the absence of transverse expansion, which is
known to be important for systems having a large multiplicity, the
life-time can become as large as several thousand fm/$c$. This would
then mimic the case of particles contained in a box and lead to
the scaling behaviour assumed generally.

\section{OUTLOOK}

In brief, the theoretical description of single photons from relativistic
heavy collisions has reached a high degree of sophistication.
Several  approaches for the evaluation of rates and the
evolution of the collision dynamics have been discussed in the
literature. The only available data, namely 
the upper limits in the sulphur induced collisions at the SPS energies,
are already indicative of unacceptability of any description
which does not involve a phase transition to QGP, unless of-course
we are comfortable with very high hadronic densities.
 Treatments invoking
a phase transition provide an agreement with the upper limit in spite
of the differences in the details of the evolution mechanism. The final
data from Pb$+$Pb collisions at SPS energies are eagerly awaited.
It has been argued that the dilepton excess measured by the CERES
experiment requires
that the mass of $\rho$ mesons reduces considerably in the dense matter
produced in such collisions. If true, this will also affect the
single photon considerably. This can be of great interest.

The truly clear signals of the production of single photons are
expected to emerge at RHIC and LHC energies, where it may become
possible to get an information about the densities of quarks and gluons 
at very early times, as the plasma may be far from equilibrium, but very
hot. In fact, it may even become possible to measure diphotons~\cite{diph}
 (from $q\overline{q} \rightarrow \gamma \gamma$) and even attempt a photon
interferometry~\cite{inter} to obtain an information about the space-time
 details of the early stages of the plasma.
 
 We also feel that the scaling behaviour (Eq.~4) seen here may be useful
 in the determination of the equation of state of the hadronic matter,
 and also in deciding the actual strength of the signal of single photons.

\section*{ACKNOWLEDGEMENTS}

This work has been done in collaboration with J. Cleymans, M. G. Mustafa,
B. M\"{u}ller, K. Redlich,
and B. Sinha. We are grateful to Terry Awes, Hans Gutbrod, Joe Kapusta,
Dipali Pal, Pradip Roy, Vesa Ruuskanen,  Sourav Sarkar,
 and Helmut Satz for many helpful discussions.

\newpage


\begin{thebibliography}{39}

\bibitem{terry} R. Albrecht et al., WA80 Collaboration,
Phys. Rev. Lett. {\bf 76} (1996) 3506; 
T. C. Awes, these proceedings.

\bibitem{axel} G. Agakichev et al., CERES Collaboration,
Phys. Rev. Lett. {\bf 75} (1995) 1272;
A. Drees, these proceedings;
C. Gale, these proceedings.


\bibitem{joe} J. Kapusta, P. Lichard, and D. Seibert, Phys. Rev.
D {\bf 44}, 2774 (1991).

\bibitem{rud1} R. Baier, H. Nakkagawa, A. Niegawa, and K. Redlich,
Phys. Rev. D {\bf 45} (1992) 4323.


\bibitem{traxler} C. T. Traxler, H. Vija, and M. H. Thoma,
Phys. Lett. B {\bf 346} (1995) 329.

\bibitem{rud2} R. Baier, M. Dirks, K. Redlich, and D. Schiff,
University of Bielefeld Preprint, 1997.

\bibitem{henning} P. Henning and E. Quack, Phys. Rev. D {\bf 54} (1996)
3125.

\bibitem{hirano} T. Hirano, S. Muroya, and M. Namiki, these proceedings.

\bibitem{roy} P. K. Roy, D. Pal, S. Sarkar, D. K. Srivastava, and
B. Sinha, Phys. Rev. C {\bf 53} (1996) 2364;
D. Pal, P. K. Roy, S. Sarkar, D. K. Srivastava, and B. Sinha,
Phys. Rev. C {\bf 55} (1997) 1467.


\bibitem{str} M. Strickland, Phys. Lett. B {\bf 331} (1994) 245.


\bibitem{li} L. Xiong, E. Shuryak, and G. E. Brown, Phys. Rev. D {\bf 46},
3798 (1992).

\bibitem{song} C. S. Song, Phys. Rev. C {\bf 47}, 2861 (1993).

\bibitem{kevin} K. Haglin, Phys. Rev. C {\bf 50} (1994) 1688.

\bibitem{kim} J. K. Kim, P. Ko, K. Y. Lee, and S. Rudaz,
Phys. Rev. D {\bf 53} (1996) 4787.

\bibitem{steel} J. V. Steele, H. Yamagishi, and I. Zahed,
Phys. Lett. B {\bf 384} (1996) 255;
 J. V. Steele, H. Yamagishi, and I. Zahed, hep-ph/9704414.
 
\bibitem{sh} E. V. Shuryak and Li Xiong, Phys. Lett. B {\bf 333} (1994) 316.

\bibitem{bj} J. D. Bjorken, Phys. Rev. D {\bf 27} (1983) 140.

\bibitem{land} L. D. Landau, Izv. Akad. Nauk. SSSR, Ser. Fiz. {\bf 17}
(1953) 51.

\bibitem{kms} J. Kapusta, L. McLerran, and D. K. Srivastava,
Phys. Lett. B {\bf 283} (1992) 145.

\bibitem{nad} H. Nadeau, Phys. Rev. D {\bf 48} (1993) 3182.


\bibitem{prd} J. Alam, D. K. Srivastava, B. Sinha, and D. N. Basu,
Phys. Rev. D {\bf 48}, 1117 (1993);
D. K. Srivastava, S. Sarkar, P. K. Roy, D. Pal, and B. Sinha,
Phys. Lett. B {\bf 379} (1996) 54.


\bibitem{arbex} N. Arbex, U. Ornik, M. Pl\"{u}mer, A. Timmermann,
and R. M. Weiner, Phys. Lett. B {\bf 354} (1995 307.

\bibitem{dum} A. Dumitru, U. Katschner, J. A. Marhun, H. St\"{o}cker,
W. Greiner, and D. H. Rischke, Phys. Rev. C {\bf 51} (1995) 2166.

\bibitem{neu} J. J. Neumann, D. Seibert, and G. Fai, Phys. Rev. C {\bf 51}
(1995) 1460.

\bibitem{lok} I. P. Lokhtin and A. M. Snigirev, Phys. Lett. B {\bf 378}
(1996) 247.

\bibitem{yu} Y. A. Tarasov, Phys. Lett. B {\bf 379} (1996) 279.

\bibitem{soll} J. Sollfrank, P. Huovinen, M. Kataja, P. V. Ruuskanen,
M. Prakash, and R. Venugopalan, Phys. Rev. C {\bf 55} (1997) 392.

\bibitem{crs1} J. Cleymans, K. Redlich, and D. K. Srivastava,
  Phys. Rev. C {\bf 55} (1997) 1431.

\bibitem{burk} B. K\"{a}mpfer and O. P. Pavlenko, Z. Phys. C {\bf 62}
(1994) 491.

\bibitem{sgm} D. K. Srivastava, M. G. Mustafa, and B. M\"{u}ller,
nucl-th/9601141.

\bibitem{prl} D. K. Srivastava and B. Sinha, Phys. Rev. Lett. {\bf 73}
(1994) 2421.

\bibitem{crs2}J. Cleymans, K. Redlich, and D. K. Srivastava,
 Preprint 1997.

\bibitem{HK} R. C. Hwa and K. Kajantie, Phys. Rev. D {\bf 32} (1985) 1109.

\bibitem{vesa} H. von Gersdorff, M. Kataja, L. McLerran, and P. V.
Ruuskanen, Phys. Rev. D {\bf34} (1986) 794.

\bibitem{johanna} P. Braun-Munzinger, J. Stachel, J. P. Wessels, and
N. Xu, Phys. Lett. B {\bf 365} (1996) 1; Phys. Lett. B {\bf 344} (1995) 43.

\bibitem{ash} A. K. Choudhuri, Phys. Rev. C {\bf 51} (1951) R2889.

\bibitem{sspc} K. J. Eskola, B. M\"{u}ller, and X. N. Wang,
Phys. Lett. B {\bf 374} (1996) 20.

\bibitem{ssg} D. K. Srivastava, B. Sinha, and C. Gale, Phys. Rev. C
{\bf 53} (1996) R567.


\bibitem{diph} D. K. Srivastava, B. Sinha, and T. Awes, Phys. Lett. B
{\bf 387} (1996) 21.

\bibitem{inter} D. K. Srivastava and J. Kapusta, Phys. Lett. B {\bf 307} 
(1993) 1;
D. K. Srivastava and J. Kapusta, Phys. Rev. C {\bf 48} (1993) 1335;
D. K. Srivastava, Phys. Rev. D {\bf 49} (1994) 4523;
D. K. Srivastava and C. Gale, Phys. Lett. B {\bf 319} (1994) 407;
D. K. Srivastava and J. Kapusta, Phys. Rev. C {\bf 50} (1994) 505.


\end{thebibliography}
\end{document}